\documentclass[aps,pra,epsfigure,twocolumn,showpacs]{revtex4}
\usepackage{dcolumn}    
\usepackage{bm} 
\usepackage{graphicx}
\usepackage{amsmath}    
\usepackage{latexsym}
\usepackage{amsfonts}   
\usepackage{amssymb}
\usepackage{array}      
\usepackage{epsfig}
\usepackage{times}
\usepackage{txfonts}
\usepackage{epstopdf}
\usepackage{pifont}
\usepackage{dsfont}
\usepackage{amscd}
\usepackage{amsfonts}
\usepackage{textcomp}

\newcommand{\ket}[1]{\left\vert#1\right\rangle}

\newcommand{\bra}[1]{\left\langle#1\right\vert}

\newcommand{\be}{\begin{equation}}
\newcommand{\ee}{\end{equation}}
\newcommand{\bea}{\begin{eqnarray}}
\newcommand{\eea}{\end{eqnarray}}

\begin{document}
\title{More nonlocality with less entanglement in a tripartite atom-optomechanical system}
\author{Jing Zhang$^1$, Tiancai Zhang$^1$, Andr\'e Xuereb$^{3,4}$, David Vitali$^2$ and Jie Li$^{2,3}$}
\affiliation{$^1$State Key Laboratory of Quantum Optics and Quantum Optics Devices, Institute of Opto-Electronics, Shanxi University, Taiyuan 030006, China \\
$^2$School of Science and Technology, Physics Division, University of Camerino, Camerino (MC), Italy \\
$^3$Centre for Theoretical Atomic, Molecular and Optical Physics, School of Mathematics and Physics, Queen's University, Belfast BT7 1NN, United Kingdom\\
$^4$Department of Physics, University of Malta, Msida MSD 2080, Malta}

\begin{abstract}
We study quantum effects in hybrid atomic optomechanics in a system comprising a cloud of atoms and a mobile mirror mediated by a single-mode cavity. Tripartite nonlocality is observed in the atom-light-mirror system, as demonstrated by the violation of the Mermin-Klyshko (MK) inequality. It has been shown [C. Genes, et al., PRA 77, 050307 (R) (2008)] that tripartite entanglement is optimized when the cavity is resonant with the anti-Stokes sideband of the driving laser and the atomic frequency matches the Stokes one. However, we show that this is not the case for the nonlocality. The MK function achieves {\it minima} when the atoms are resonant with both the Stokes and anti-Stokes sidebands, and unexpectedly, we find violation of the MK inequality only in a parameter region where entanglement is far from being maximum. A negative relation exists between nonlocality and entanglement with consideration of the possibility of bipartite nonlocality in the violation of the MK inequality. We also study the non-classicality of the mirror by post-selected measurements, e.g. Geiger-like detection, on the cavity and/or the atoms. We show that with feasible parameters Geiger-like detection on the atoms can effectively induce mechanical non-classicality.
\end{abstract}
\date{\today}
\pacs{} 
\maketitle

The lack of observation of quantum effects at the macroscopic scale reinforces the conjecture that macroscopic objects are governed by classical physics, while the microscopic world is ruled by quantum mechanics. However, quantum mechanics intrinsically shows no limitation to describe large-scale/massive systems~\cite{leggett}. Preparing macroscopic quantum states is of vital importance for understanding fundamental issues in quantum mechanics, such as decoherence and the quantum-to-classical transition~\cite{zurek}, collapse models of the wave function~\cite{collapse}, and so on. Optomechanics, addressing the coupling of optical and mechanical degrees of freedom via radiation pressure~\cite{optomreview}, provides an ideal platform to generate and control quantum mesoscopic/macroscopic states of mechanical systems thanks to its intrinsic nonlinear light-matter interactions.

Over the past few years, successful advances in nano- and micro-mechanical engineering, in particular mechanical oscillators cooled into (or close to) their ground state~\cite{Teufel,Chancaltech}, have made it possible to prepare mechanical quantum states. Preparing quantum states either for the light mode or the mechanical oscillator is a fascinating (though challenging) goal in the field of optomechanics~\cite{nonclassical,sbose}. Nonclassical mechanical states can be generated by the optomechanical nonlinearity intrinsic in the strong coupling regime~\cite{nunnenkamp,qian}, by injecting squeezed light into the cavity (the squeezing is thus transferred from light to the mechanical degree of freedom)~\cite{Jahne,huang}, by post-selected measurements on the optical field~\cite{jiesnjp,milburnnjp}, and so on. 

Recently, it has been reported that hybrid atom-assisted optomechanics shows advantages in many aspects~\cite{hybridreview}. To name but a few, atoms induce an additional nonlinear effect, which enhances the optomechanical interaction and, moreover, results in a squeezed state of the mechanical mode~\cite{nori}; atoms boost the cooling of the mechanical motion~\cite{atomcooling} and can be utilized to prepare non-Gaussian mechanical states~\cite{cirac2013}; and the strong coupling between an atom/atoms and a mechanical oscillator allows to realize quantum control of the oscillator via manipulating the atom/atoms~\cite{atomOM}. It has also been shown that a genuine tripartite entangled stationary state of an atom-cavity-mirror system can be produced~\cite{vitali3,ben}. Very recently, nonlocality of an atom and a mechanical oscillator mediated by a single-mode cavity has been studied~\cite{vlatko}.

Being incompatible with local realism, nonlocality is demonstrated by the violation of Bell inequalities~\cite{bell}, and witnesses a type of quantum correlations that is distinct from entanglement and discord. It has been proven that a {\it pure} bipartite/tripartite entangled state is nonlocal~\cite{ennonlocal}. However, in general, nonlocality and entanglement are different properties: a system in an entangled {\it mixed} state does not necessarily possess nonlocal correlations~\cite{Werner1989}. The relationship between both is much more subtle for {\it mixed} states and far from being clear especially in multipartite cases~\cite{nonlocalRMP}. It is thus of fundamental importance to explore their relationship in such states, practicably starting from a specific case, which is the main theme of this paper. Multipartite nonlocality has been investigated in continuous variable (CV) systems~\cite{cvnonlocal}. In such cases, it is usually demonstrated by the violation of Bell-like inequalities in phase space due to the systems' infinite-dimensional Hilbert spaces~\cite{phasespace}.

In this paper, we investigate tripartite nonlocality and entanglement in a hybrid optomechanical system composed of an atomic ensemble placed within an optomechanical cavity. The system is subject to noise and dissipation, and its state is in general highly mixed. It therefore represents an ideal platform for exploring the relationship between nonlocality and entanglement in mixed states. It has been reported~\cite{vitali3} that robust genuine tripartite entanglement can be generated using experimentally feasible parameters under the following conditions: (i) the cavity is resonant with the anti-Stokes sideband of the driving laser (i.e. the mechanical cooling regime); (ii) the atomic frequency matches the Stokes sideband; (iii) the effective optomechanical/atom-light coupling is large compared to cavity/atomic decay. Under these conditions, however, we show that the value of Mermin-Klyshko (MK) function~\cite{MKineq} (from which the MK inequality is constructed whose violation denotes tripartite nonlocality) shows a negative relation with the entanglement for a wide range of the cavity decay: the MK value decreases for increasing entanglement. By relaxing the optomechanical/atom-light coupling thus diminishing the entanglement, we observed a violation of the MK inequality demonstrating nonlocal correlations shared among the atom-light-mirror system. Furthermore, we also show that the MK function achieves {\it minima} when the atoms are resonant with both the Stokes and anti-Stokes sidebands of the laser, while the former is the condition under which the tripartite entanglement is maximized.

Another theme of this paper is to explore the impact of the atomic ensemble on the non-classicality of the mechanical mode. We show that by post-selected measurements, e.g. Geiger-like detection, on the atomic state, a negative Wigner function of the mirror is observed witnessing the quantum nature of its motional state. Larger coupling strength of the system leads to a more nonclassical mechanical state.

%The remainder of the paper is structured as follows. In Sec.~\ref{system} we introduce the system and provide the details of its solution, leading toward the determination of the covariance matrix (CM) that fully describes the system. Sec.~\ref{3nonlocal} is devoted to the study of tripartite nonlocality of the system. We show that the MK inequality is violated with feasible parameters. The nonlocality is compared in detail with the tripartite entanglement of the system. In Sec.~\ref{noncalssical} we study the non-classicality of the mechanical subsystem by performing post-selected measurements on the cavity and/or the atoms. Finally, we conclude in Sec.~\ref{concl}.

\section{The system} 
\label{system}

We consider a pump laser at frequency $\omega_l$ driving a Fabry-Perot cavity with a light vibrating end mirror of mass $m$ and mechanical frequency $\omega_m$. An ensemble of $N$ two-level atoms with natural frequency $\omega_a$ is placed inside the cavity. In a unitary picture, without considering any dissipation and decoherence, the Hamiltonian of the system is
\begin{equation}
\begin{split}
H&{=}\hbar \omega_c c^\dagger c+\frac{\hbar}{2}\omega_a S^z+\frac{\hbar}{2}\omega_m (q^2+p^2)\\
&-\hbar\chi c^\dagger c q+\hbar g(S^+c+S^-c^\dagger)+i\hbar\varepsilon(c^\dagger e^{-i\omega_lt}-ce^{i\omega_lt}),
\end{split}
\label{hamil}
\end{equation}
where $\omega_c$ is the cavity frequency, and $c$ ($c^\dagger$) the corresponding optical annihilation (creation) operators. $q$ and $p$ are the dimensionless mechanical mode position-like and momentum-like operators, and $S^{\pm}$ are the collective spin operators of the ensemble of atoms defined as $S^{\pm,z}{=}\sum_i\sigma^{\pm,z}_i$ ($i{=}1,2,...,N$) with Pauli matrices $\sigma^{\pm}$ and $\sigma^{z}$, which satisfy the commutation relations $[S^+,S^-]{=}S^z$ and $[S^z,S^{\pm}]=\pm2S^{\pm}$. $\chi$ and $g$ are the optomechanical and atom-cavity coupling given by $\chi=(\omega_c/L)\sqrt{\hbar/m\omega_m}$ with $L$ the cavity length, and $g=d\sqrt{\omega_c/2\hbar\epsilon_0 V}$ with $d$ the dipole moment of the atomic transition, $\epsilon_0$ the vacuum permittivity and $V$ the volume of the cavity mode. $\varepsilon$ is the coupling between the driving laser and the cavity field, which is related to the pump power $P$ and the cavity decay $\kappa$ by $\varepsilon=\sqrt{2P\kappa/\hbar\omega_l}$.

The dynamics of this tripartite system is in principle complicated. For simplifying the calculation, we adopt the treatment employed in Ref.~\cite{vitali3}, by assuming the low atomic excitation limit, i.e. atoms are initially set in the ground state, and the excitation probability of a single atom is small. In such a case, the dynamics of the atomic polarization can be described by the bosonic annihilation operator $a=S^-/\sqrt{|\langle S^z\rangle|}$ and its Hermitian conjugate $a^\dagger$, which satisfy the commutation relation $[a,a^\dagger]{=}1$. In a rotating frame at the pump frequency $\omega_l$, the dynamics of such a system can then be described by a set of quantum Langevin equations as
\begin{equation}
\begin{split}
\dot{q}&=\omega_m p,\\
\dot{p}&=-\omega_m q-\gamma_m p+\chi c^\dagger c+\xi,\\
\dot{c}&=-(\kappa+i\Delta_c)c+i\chi cq-ig_Na+\varepsilon+\sqrt{2\kappa}c_{in},\\
\dot{a}&=-(\gamma_a+i\Delta_a)a-ig_Nc+\sqrt{2\gamma_a}a_{in},\\
\end{split}
\end{equation}
where $\gamma_m$ ($\gamma_a$) is the decay rate of the mechanical oscillator (atoms), and $\Delta_c{=}\omega_c{-}\omega_l$ ($\Delta_a{=}\omega_a{-}\omega_l$) is the cavity (atomic) detuning with respect to the laser. $g_N{=}\sqrt{N}g$ is the coupling between the cavity and the collective atomic mode.
$\{\xi,c_{in},a_{in}\}$ are the noise operators of the system affecting the mirror, optical and atomic field, respectively. The Langevin force operator $\xi$, which accounts for the Brownian motion of the mirror, is auto-correlated as~\cite{DVitali}
\begin{equation}
\langle\xi(t)\xi(t')\rangle{=}\frac{\gamma_m}{2\pi\omega_m}\int\omega \text{e}^{-i\omega(t-t')}[\text{coth}(\frac{\hbar\omega}{2k_BT})+1]\text{d}\omega,
\end{equation}
with $k_B$ the Boltzmann constant, $T$ the temperature of the phononic environment. For a large mechanical quality factor, the above correlation function reduces to a $\delta$ function~\cite{deltafunction}. When the cavity and atomic modes are prepared in coherent states, the only nonzero correlations of $c_{in}$ and $a_{in}$ are $\langle c_{in}(t)c_{in}^{\dag}(t')\rangle{=}\langle a_{in}(t)a^{\dag}_{in}(t')\rangle{=}\delta(t-t')$~\cite{noisebook}.

In what follows, to enhance the optomechanical coupling, we assume the cavity is strongly pumped, i.e. $|\alpha_s|\gg 1$, where $\alpha_s$ is the amplitude of the steady-state cavity field, which can be acquired by solving the nonlinear equation $\alpha_s[\kappa+i\Delta_c-i\chi^2|\alpha_s|^2/\omega_m+g^2_N/(\gamma_a+i\Delta_a)]=\varepsilon$. In that case, one can then approximate the quadrature operators of the system $O{=}(q,p,X,Y,x,y)$ as $O_i{\simeq}\langle O_i\rangle{+}\delta O_i$, with $\langle O_i\rangle$ the `large' mean value of each operator and $\delta O_i$ the corresponding `small' fluctuation, where we introduced $X{=}(c^\dag{+}c)/\sqrt2$, $Y{=}i(c^\dag{-}c)/\sqrt2$, and $x{=}(a^\dag{+}a)/\sqrt2$, $y{=}i(a^\dag{-}a)/\sqrt2$ the position- and momentum-like operators of the optical and atomic modes, respectively. In such a way, the dynamics of the system takes a linear form that simplifies the cumbersome calculation. The resulting dynamics of the fluctuation operators $\delta O{=}(\delta q,\delta p,\delta X,\delta Y,\delta x,\delta y)$ is described by a set of Langevin equations
\begin{equation}
\begin{split}
\delta\dot{q}&=\omega_m\delta p, \\
\delta\dot{p}&=-\omega_m\delta q-\gamma_m\delta p+\chi_{e\!f\!f}\delta X+\xi,  \\
\delta\dot{X}&=-\kappa\delta X+\tilde{\Delta}_c\delta Y+g_N\delta y+\sqrt{2\kappa} X_{in},  \\
\delta\dot{Y}&=-\kappa\delta Y-\tilde{\Delta}_c\delta X+\chi_{e\!f\!f}\delta q-g_N\delta x +\sqrt{2\kappa} Y_{in},  \\
\delta\dot{x}&=-\gamma_a\delta x+\Delta_a\delta y+g_N\delta Y+\sqrt{2\gamma_a} x_{in},  \\
\delta\dot{y}&=-\gamma_a\delta y-\Delta_a\delta x-g_N\delta X+\sqrt{2\gamma_a} y_{in},  \\
\end{split}
\label{QLE}
\end{equation}
with the effective optomechanical coupling $\chi_{e\!f\!f}{=}\sqrt{2}\chi\alpha_s$ (without losing generality, we have taken $\alpha_s$ as real), the effective cavity detuning $\tilde{\Delta}_c=\Delta_c-\chi^2_{e\!f\!f}/2\omega_m$, and noise operators $X_{in}{=}(c^\dag_{in}{+}c_{in})/\sqrt2$, $Y_{in}{=}i(c^\dag_{in}{-}c_{in})/\sqrt2$, and $x_{in}{=}(a^\dag_{in}{+}a_{in})/\sqrt2$, $y_{in}{=}i(a^\dag_{in}{-}a_{in})/\sqrt2$. Equations~(\ref{QLE}) can be solved directly in the frequency domain by taking the Fourier transform of each equation above. The correlation function of any pair of fluctuation operators is then acquired as
\begin{equation}
V_{ij}=\frac{1}{4\pi^2}\iint\!\text{d}\omega\text{d}\Omega e^{-i(\omega+\Omega)t}V_{ij}(\omega,\Omega),
\label{correlations}
\end{equation}
where $V_{ij}(\omega,\Omega)=\langle\{v_i(\omega),v_j(\Omega)\}\rangle/2 ~(i,j\,{=}\,1,..,6)$ is the correlation function between elements $i$ and $j$ of $v(\omega)=(\delta q(\omega), \delta p(\omega),\delta X(\omega),\delta Y(\omega),\delta x(\omega),\delta y(\omega) )$. All the elements of $V_{ij}(\omega,\Omega)$ constitute a $6\times6$ covariance matrix (CM) of the system in the frequency domain. $V_{ij}(\omega,\Omega)$ contains a delta function $\delta(\omega+\Omega)$, which leads to the disappearance of $e^{-i(\omega+\Omega)t}$ in Eq.~\eqref{correlations} after the integrations. As this $\delta$ function is a consequence of the stationarity of the noises~\cite{steadyCM}, the resulting time-independent CM ${\bm \sigma}$ with elements defined in Eq.~\eqref{correlations} describes the steady state of the system. Our hybrid optomechanical system is fully determined by the CM ${\bm \sigma}$. Being a physical state, this CM should satisfy the Heisenberg-Robertson uncertainty principle ${\bm\sigma}+i\Omega_3/2\ge0$~\cite{SimonMukunda} with $\Omega_3{=}\oplus^3_{j=1}i\sigma_y$ the so-called symplectic matrix and $\sigma_y$ the $y$-Pauli matrix. Note that, owing to the linearization of the dynamics and the fact that all noises are Gaussian, the dynamical map of the system preserves the Gaussian nature of any input state.

\begin{figure*}[t]
{\bf (a)}\hskip4.3cm{\bf (b)}\hskip4.2cm{\bf (c)}\hskip4.2cm{\bf (d)}
\includegraphics[width=0.24\linewidth]{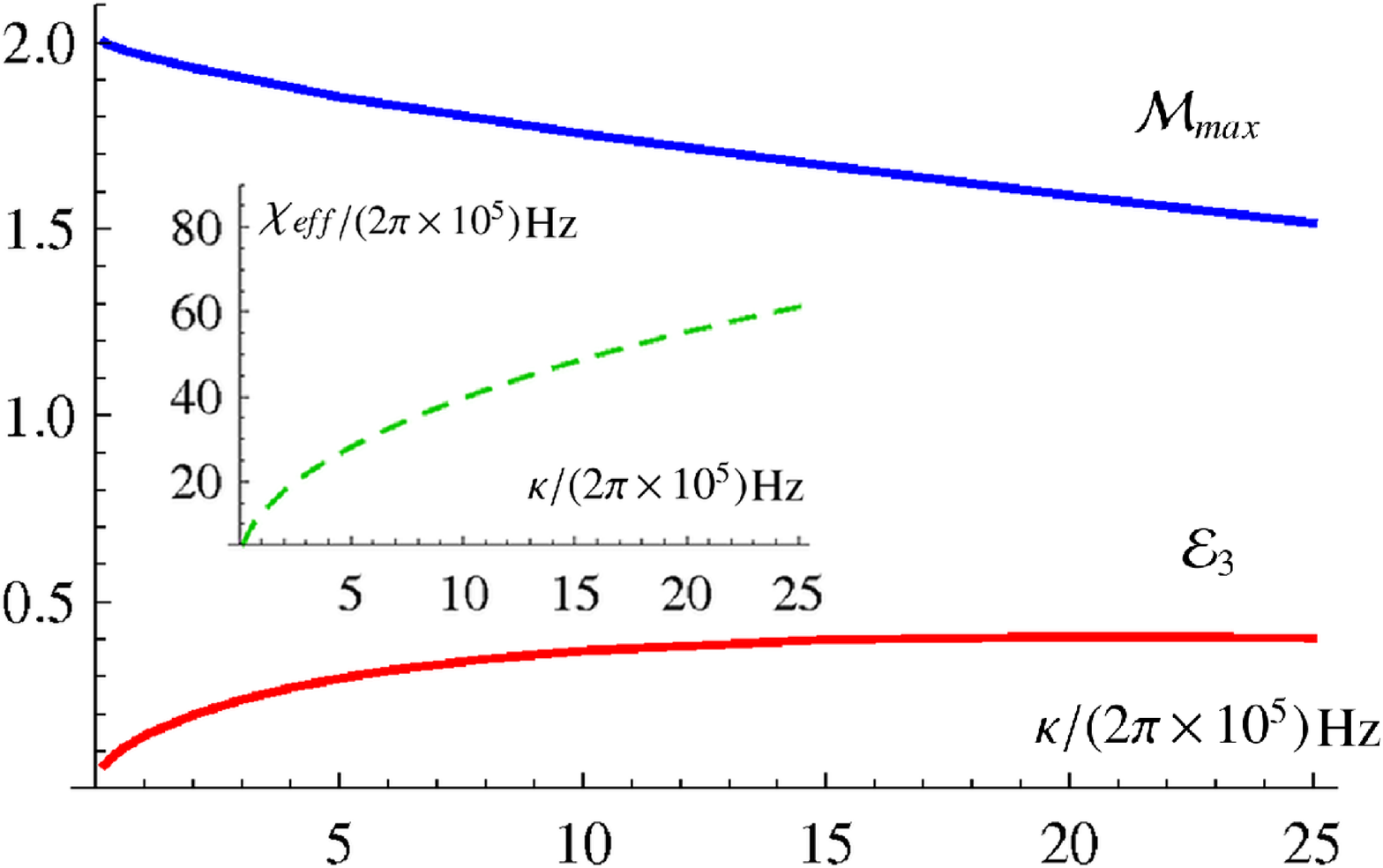}~\includegraphics[width=0.255\linewidth]{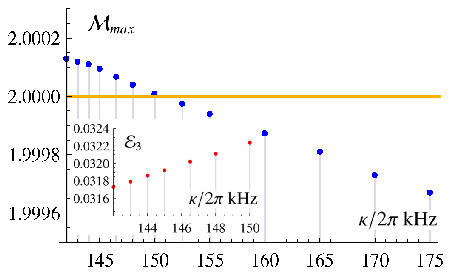}~~\includegraphics[width=0.26\linewidth]{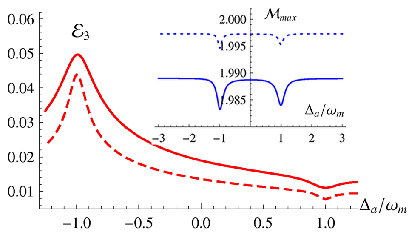}~~\includegraphics[width=0.23\linewidth]{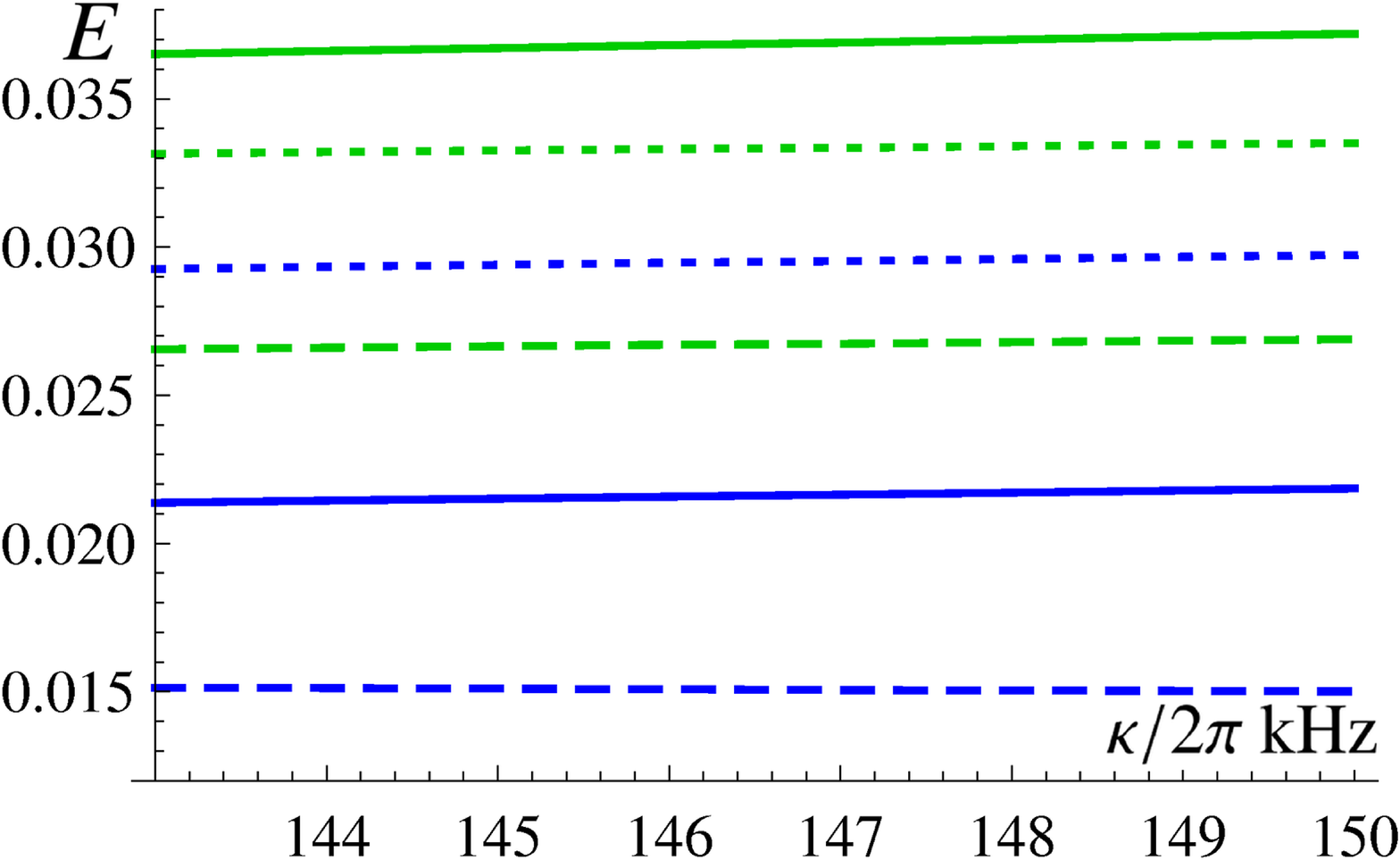}
\caption{{\bf (a)} Tripartite entanglement ${\cal E}_3$ and MK function ${\cal M}_{max}$ versus cavity decay $\kappa$. Inset shows the steady-state coupling $\chi_{e\!f\!f}$ as a function of $\kappa$. Detuning is optimized for ${\cal E}_3$: $\tilde{\Delta}_c{=}\omega_m$ and $\Delta_a{=}-\omega_m$. Cavity length is taken as $L{=}1$ mm. {\bf (b)} Tripartite nonlocality demonstrated by the violation of the MK inequality (inset for the entanglement ${\cal E}_3$) versus cavity decay $\kappa$. Parameters are taken same values as in {\bf (a)} but for a larger cavity $L{=}5$ mm.  {\bf (c)} Tripartite entanglement ${\cal E}_3$ (red) and MK function ${\cal M}_{max}$ (blue) versus atomic detuning $\Delta_a$. The same conditions as in {\bf (b)} but for two cases of cavity decay $\kappa/2\pi{=}10^6$ Hz (solid), $\kappa/2\pi{=}5\times10^5$ Hz (dashed). ${\cal E}_3$ takes a maximum at $\Delta_a{=}-\omega_m$, while ${\cal M}_{max}$ achieves {\it minima} when $\Delta_a{=}\pm\omega_m$. {\bf (d)} Bipartite entanglement $E_{ij}$ (blue) and $E_{i|jk}$ (green) as a function of $\kappa$: $E_{12}$, $E_{1|23}$ (solid); $E_{23}$, $E_{2|13}$ (dashed); $E_{13}$, $E_{3|12}$ (dotted). The same conditions as in {\bf (b)}.}
\label{enMK}
\end{figure*}

\section{Tripartite nonlocality versus tripartite entanglement} 
\label{3nonlocal}

In this Section, we devote ourselves to studying nonlocal properties of our hybrid tripartite system and, moreover, exploring the relationship between nonlocality and entanglement, which are two central concepts in quantum physics. Up to now, their relationship is far from being answered especially in multipartite {\it mixed} states~\cite{nonlocalRMP}. It should be pointed out that the definition of tripartite/multipartite nonlocality is still not clear and unified~\cite{nonlocalRMP,GisinBell}. Usually, it is defined as {\it genuine} (or three-way) tripartite/multipartite nonlocality denoted by the violation of Svetlichny inequality (SI)~\cite{Svet}. However, there exists tripartite nonlocality (here we mean nonlocality involving all three parties) that does not violate SI, i.e., is not necessarily {\it genuine}: for example, in a three-particle system, particle 1 is nonlocally correlated with the other two, while particles 2 and 3 are locally correlated. In such a case, nonlocal correlations could be signalled by the violation of other weaker Bell-like inequalities, e.g. the MK inequality. The `drawback' of the MK is that its violation admits states with only two-particle nonlocal correlations present~\cite{Cereceda}. It is in principle impossible to distinguish tripartite nonlocality from bipartite ones in the violation of MK inequality. In what follows, we will focus on testing the MK inequality in our highly mixed three-mode Gaussian state in view of the failure of violating the SI in such a state. It should be pointed out that the violation of SI is rather demanding, with genuinely multipartite entangled states of the $W$ form achieving values of the Svetlichny function only slightly larger than 4 \cite{Cereceda}, and for Gaussian states, violation of the SI seems impossible when the purity falls below 0.86 \cite{AdessoPRL}.

Given the CM of the system, one can write its characteristic function $\zeta(O)={\rm exp}(-O{\bm \sigma}O^{\rm T})$~\cite{parisbook}. The Wigner function is defined as the Fourier transform of $\zeta(O)$. For our zero-mean three-mode Gausssian state, the Wigner function is given by
\begin{equation}
W_{\sigma}(O)=\frac{{\rm exp}(-O{\bm \sigma}^{-1}O^{\rm T})}{\pi^3 \sqrt{{\rm det}[{\bm \sigma}]}},
\end{equation}
where $O$ denotes the phase-space variables associated with the fluctuation operators $\delta O$. Nonlocality of CV systems can be tested in the phase space by adopting the displaced parity operator $\Pi(\lambda)=D(\lambda)\,\Pi\,D^{\dagger}(\lambda)$ to be measured on each mode~\cite{Wodkiewicz}, with the Weyl displacement operator $D(\lambda){=}{\rm exp}(\lambda b^{\dagger}{-}\lambda^* b)$ ($\lambda\in\mathbb{C}$) and the parity operator
\begin{equation}
\Pi=(-1)^n=\sum^\infty_{n{=}0}(\ket{2n}\bra{2n}-\ket{2n+1}\bra{2n+1}),
\end{equation}
where $n{=}b^{\dagger}b$ is the bosonic number operator and $\ket{n}$ the $n$-excitation Fock state. The key of such a phase-space approach is that the mean value of the displaced parity operator is connected to the Wigner function, i.e. $\langle\Pi(\lambda)\rangle{=}(\pi/2)W(\lambda)$~\cite{Wodkiewicz}. Consequently, for our three-mode Gaussian system, the MK function can be rewritten in the phase space as
\begin{equation}
\begin{split}
{\cal M}_3&=\frac{\pi^3}{8}[W_{\sigma}(O'_1,O_2,O_3)+W_{\sigma}(O_1,O'_2,O_3) \\
&+W_{\sigma}(O_1,O_2,O'_3)-W_{\sigma}(O'_1,O'_2,O'_3)],
\end{split}
\label{mk3}
\end{equation}
where $O_1{=}\{q,p\}$, $O_2{=}\{X,Y\}$ and $O_3{=}\{x,y\}$ that fully describe the mirror, cavity, and atoms subsystems, respectively, and $O'_i$ embodies different values of the same quadrature operators of $O_i$. Any local realistic theory imposes the bound $|{\cal M}_3|\le 2$. Bipartite and/or tripartite nonlocal correlations among the system result in a violation of the MK inequality, i.e., $|{\cal M}_3|>2$. In the following, we define ${\cal M}_{max}$ as the maximum of ${\cal M}_3$ optimized over the full range of $\{q,p,X,Y,x,y,q',p',X',Y',x',y'\}$. 

In order to study the relation between nonlocality and entanglement in our mixed three-mode system, we introduce the genuine tripartite entanglement, which can be quantified by tripartite negativity~\cite{3entangle}, defined as
\begin{equation}
{\cal E}_3=(E_{1|23}\,E_{2|13}\, E_{3|12})^{1/3},
\end{equation}
where $E_{i|jk}$ is the one-vs-two-mode entanglement between mode $i$ and modes $j+k$ ($i,j,k=1,2,3$). When $E_{i|jk}>0~(\forall~i,j,k=1,2,3)$, i.e. all one-vs-two-mode bipartitions in the system are inseparable, the tripartite negativity ${\cal E}_3>0$ implies the existence of genuine
tripartite entanglement shared within the system~\cite{Giedke}. To quantify $E_{i|jk}$, we employ the logarithmic negativity~\cite{logneg}, which is calculated as $E_{i|jk}=\max[0,-\ln2\tilde\nu_-]$, with $\tilde\nu_-=\min{\rm eig}|i\Omega_3(P_{i|jk}{\bm\sigma}P_{i|jk})|$, where $P_{i|jk}$ is the matrix that inverts the sign of momentum of mode $i$. Similarly, one can obtain all the bipartite entanglement $E_{ij}$ $(i,j=1,2,3)$.

The numerical results of the tripartite entanglement and nonlocality are shown in Fig.~\ref{enMK}, in which we employed the following parameters~\cite{Chancaltech,Painter}: the mass of the mirror $m{=}10$ ng, with $\omega_m/2\pi{=}10^7$ Hz, $\gamma_m/2\pi{=}100$ Hz, and phononic temperature $T=0.1$ mK; pump power $P{=}35$ mW at $\lambda_l{=}1064$ nm, and cavity length $L{=}1$ mm in Fig.~\ref{enMK} {\bf (a)}, and $L{=}5$ mm in {\bf (b)} and {\bf (c)}. In the following, we set equal optomechanical and atoms-light coupling, $\chi_{e\!f\!f}=g_N$~\cite{notecoupling} (the strength of $g_N$ can be adjusted by changing the number of atoms), and equal cavity and atomic decay, $\kappa=\gamma_a$. The tripartite entanglement can only be present within a high-finesse cavity, and a large $\chi_{e\!f\!f}$ ($g_N$) compared to decay $\kappa$ ($\gamma_a$). This could be achieved with $L{=}1$ mm and large finesse $F>10^4$. The entanglement is optimized for $\tilde{\Delta}_c{=}\omega_m$ and $\Delta_a{=}-\omega_m$~\cite{vitali3}, i.e. the cavity is resonant with the anti-Stokes sideband of the laser, while the atoms are resonant with the Stokes sideband. 

In Fig.~\ref{enMK} {\bf (a)}, we plot ${\cal M}_{max}$ and ${\cal E}_3$ for a wide range of cavity decay $\kappa$ under the above conditions. As shown, a considerable degree of ${\cal E}_3$ emerges, and, moreover, ${\cal E}_3$ is robust against the temperature surviving up to $T=15$ K for cavity finesse $F{=}3\times10^4$. As $\kappa$ becomes larger, the steady-state coupling $\chi_{e\!f\!f}$ increases remarkably (see the inset) resulting in a rising ${\cal E}_3$ till to a saturated value. Surprisingly, ${\cal M}_{max}$ declines as ${\cal E}_3$ increases. For the whole range of $\kappa$, we have not witnessed any violation of the MK inequality. Given the negative relation between ${\cal M}_{max}$ and ${\cal E}_3$, one would expect to see ${\cal M}_{max}{>}2$ under conditions where ${\cal E}_3$ is smaller. In what follows, we will show this is indeed the case. By relaxing the coupling $\chi_{e\!f\!f}$, realized by increasing the cavity length to $L$=$5$ mm, we observed a weak violation of the MK inequality when $\kappa$ takes small values (smaller $\kappa$ will not satisfy the conditon for CM ${\bm \sigma}$ being physical), as shown in Fig.~\ref{enMK} {\bf (b)}. In such a case, ${\cal E}_3$ can only achieve its maximum around 0.05 due to a much weaker $\chi_{e\!f\!f}$ [about 0.17 $\chi_{e\!f\!f}$ of getting maximum ${\cal E}_3$ for $L{=}1$ mm in Fig.~\ref{enMK} {\bf (a)}]. Unlike the robustness of entanglement towards the temperature, the nonlocality is quite fragile: ${\cal M}_{max}$ drops below 2 when $T$ rises up to 1 mK. Such a feature has also been observed in the tripartite nonlocality of the vibrational modes of trapped ions~\cite{jie3ions}. From an experimental perspective, we see that $T{\sim}0.1$ mK can not be reached with standard dilution refrigerators (which typically reaches 10 mK and hardly below). However such temperatures could be reached by employing advanced techniques such as adiabatic nuclear demagnetization refrigerators~\cite{Nguyen}. Alternatively, one could think of using GHz oscillators, for which the nonlocality properties discussed here would be visible at higher temperatures.

This negative relation between nonlocality and entanglement is confirmed by Fig.~\ref{enMK} {\bf (c)} that shows ${\cal E}_3$ and ${\cal M}_{max}$ as a function of atomic detuning $\Delta_a$, for $L$=$5$ mm, $\tilde{\Delta}_c{=}\omega_m$, and two working points of $\kappa$. Unambiguously, ${\cal E}_3$ reaches a peak as $\Delta_a{=}-\omega_m$, while ${\cal M}_{max}$ gets two {\it minima} when $\Delta_a{=}\pm\omega_m$. %It should be noted that, for smaller $\kappa$ two tiny {\it minima} are also found for ${\cal M}_{max}{>}2$, i.e. in the nonlocal regime. However, they are too inconspicuous to show. 
Since ${\cal M}_{max}$ is not sensitive to $\Delta_a$ when $\kappa$ is small, in Fig.~\ref{enMK} {\bf (b)} we used $\Delta_a{=}-\omega_m$. Evidently, as $\Delta_a$ takes values away from $\pm\omega_m$, a slight rise of ${\cal M}_{max}$ would occur. 

Since the violation of MK inequality admits bipartite nonlocal correlations~\cite{Cereceda}, we now prove the negative relation still holds even in this situation. Above all, it is necessary to specify unambiguously the {\it positive relation} between nonlocality and entanglement. Despite specific states studied, measures of the entanglement and Bell inequalities adopted, the {\it positive relation} contains two apparent meanings (not necessarily complete): (i) as the entanglement increases/decreases, the nonlocality also increases/decreases, and vice versa; (ii) for a multipartite nonlocal state with two of the bipartition entanglement $E_{A|B}^1>E_{A|B}^2$ (where A and B are subsystems regardless of the number of particles/modes comprised in both), $E_{A|B}^1$ is more likely than $E_{A|B}^2$ to violate the Bell inequality, or contributes more to the violation. Now {\it suppose the positive relation is valid} in our case, then the decreasing MK function might result from one or more declining bipartite entanglement $E_{ij}$ and/or $E_{i|jk}$, though the tripartite entanglement ${\cal E}_3$ is increasing. To ascertain this, we show in Fig.~\ref{enMK} {\bf (d)} all bipartite entanglement $E_{ij}$ and $E_{i|jk}$ at the range of $\kappa$ when ${\cal M}_{max}{>}2$ in Fig.~\ref{enMK} {\bf (b)}. It shows that only atoms-light entanglement $E_{23}$ is slightly decreasing as $\kappa$ grows. According to (i), only $E_{23}$ could lead to the decreasing ${\cal M}_{max}$, and then in view of (ii), $E_{23}$ should be larger than any other entanglement plotted in Fig.~\ref{enMK} {\bf (d)}. This is clearly against the fact that $E_{23}$ is the minimal entanglement and has the least possibility to get the MK inequality violated. Therefore, the previous assumption does not hold and thus the nonlocality and entanglement show a negative relation even if bipartite nonlocality is present in the violation of the MK inequality. The breaking of the positive relation (i) has been reported in Refs.~\cite{AAcin,cabello}.

From the above analyses, we remark that the emergence of both tripartite entanglement and nonlocality within the system requires a high-finesse cavity, a high mechanical Q factor, strong coupling $\chi_{e\!f\!f}$ ($g_N$) with relatively low decay $\kappa$ ($\gamma_a$), and the cavity is resonant with the anti-Stokes sideband. Differently, the appearance of entanglement mainly depends on the coupling strength, while nonlocality is mainly sensitive to the value of the decay rate, as illustrated in Fig.~\ref{enMK} {\bf (b)}, ${\cal M}_{max}>2$ can only exist at extra-low cavity and atomic decay. This inconsistency could be the physical reason that leads to the negative relation between both. %Of course, a deeper reason for this inconsistency must be due to the different definitions of nonlocality and entanglement. 
It should be pointed out that it is in principle hard to give a {\it general} conclusion on the relationship between both in view of various measures or Bell inequalities for the nonlocality, especially for multipartite cases which display a more complex structure than the bipartite cases~\cite{nonlocalRMP}. Conclusions may vary significantly and even become completely opposite depending on what kind of measures one adopts, specific states one studies, and whether the system is in a pure or mixed state. To be specific, by adopting different measures, Vallone {\it et al.}~\cite{cabello} find entanglement and nonlocality are inversely related for pure two-qubit/qutrit states. On the contrary, Adesso {\it et al.}~\cite{AdessoPRL} show a good agreement between both for {\it pure} three-mode Gaussian states. Nevertheless, in our {\it mixed} three-mode Gaussian states a negative relation is observed. The relationship between both has been rarely explored in multipartite {\it mixed} states~\cite{nonlocalRMP}. Our work, to the best of our knowledge, for the first time provides a concrete demonstration of the negative relation between both in such states.

\begin{figure}[t]
\hskip0.5cm{\bf (a)}\hskip2.6cm{\bf (b)}\hskip2.6cm{\bf (c)}
\includegraphics[width=0.33\linewidth]{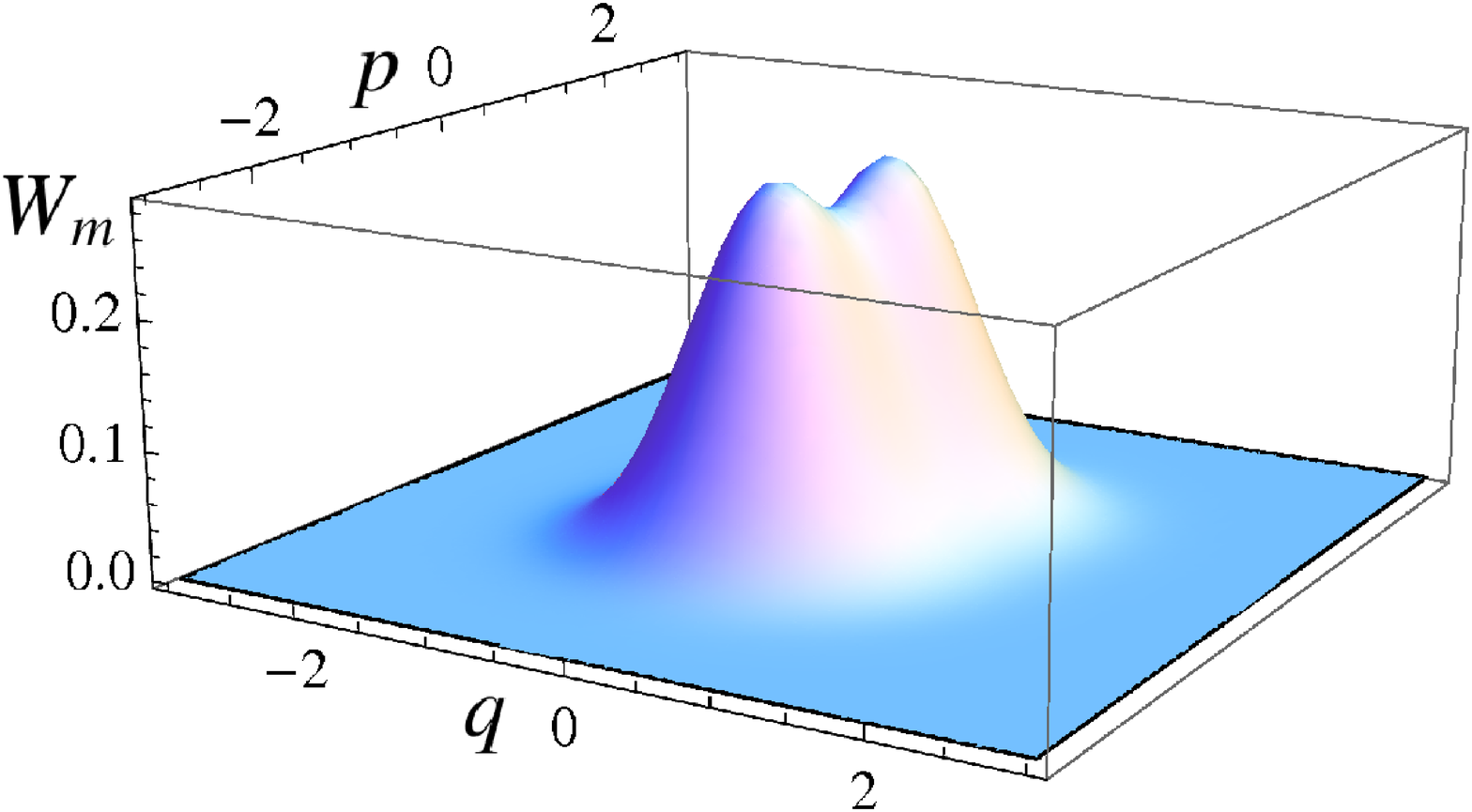}~~~\includegraphics[width=0.34\linewidth]{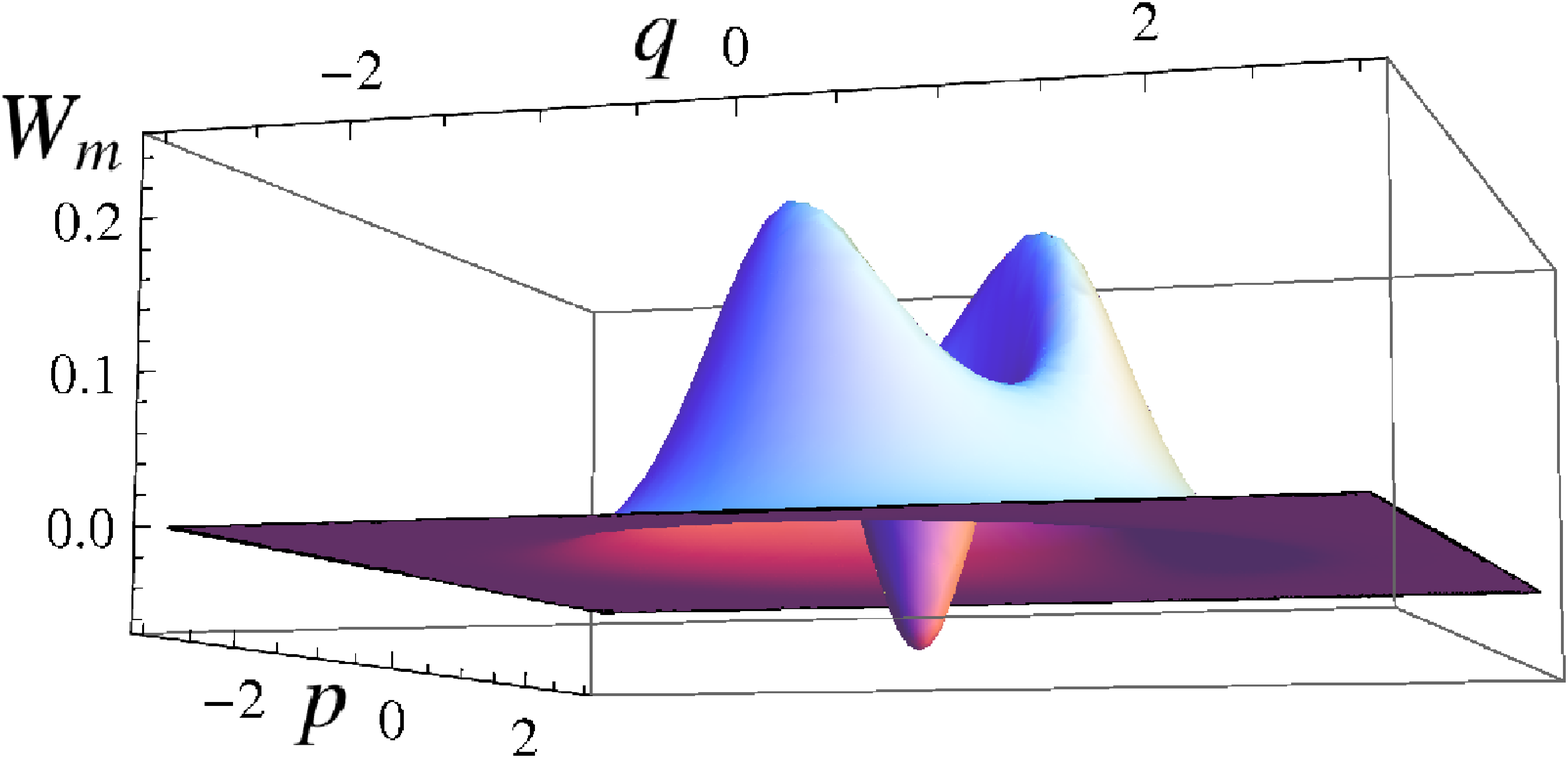}~~~\includegraphics[width=0.32\linewidth]{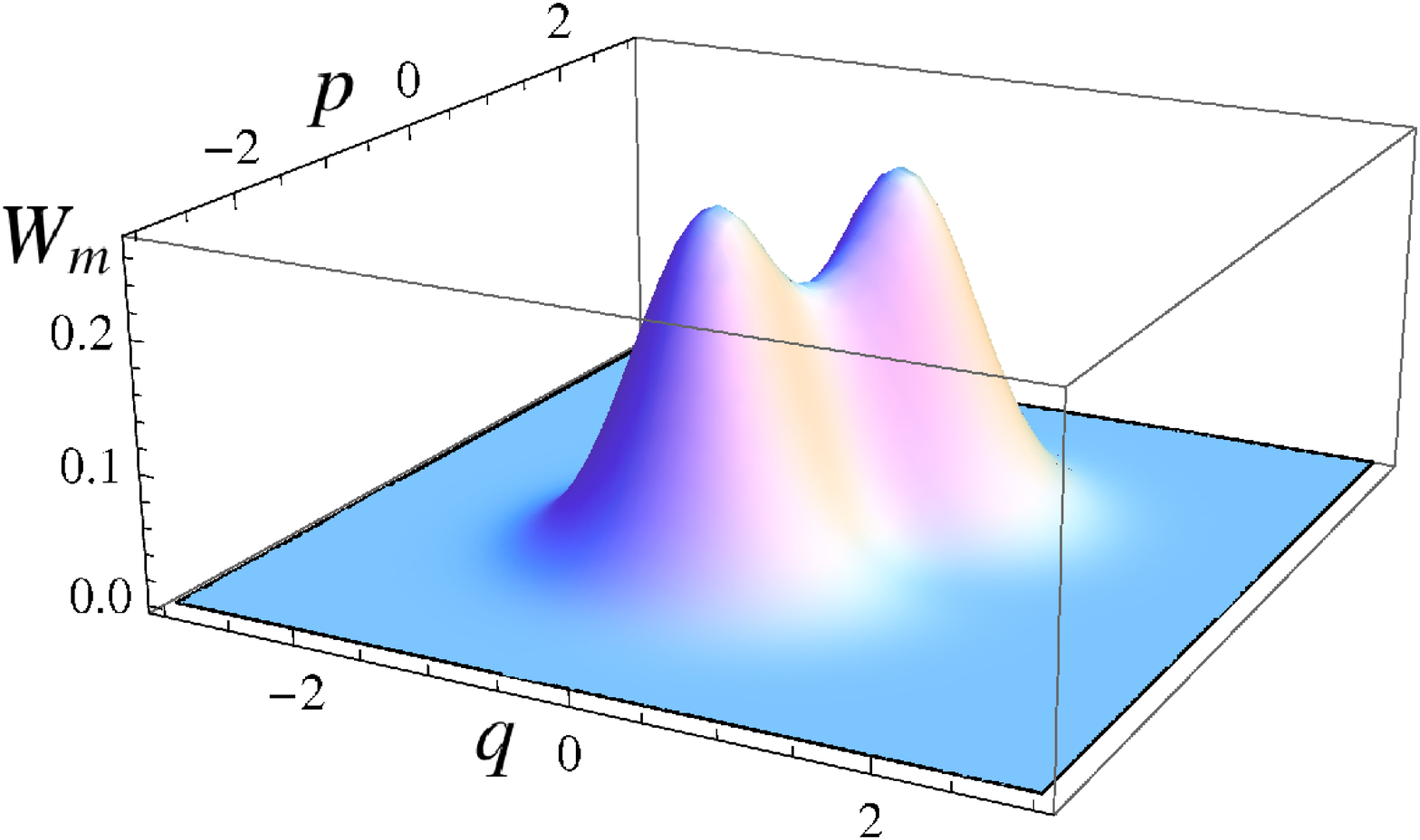}
\caption{Wigner function of the mirror with Geiger-like detection {\bf (a)} on the cavity; {\bf (b)} on the atomic state; {\bf (c)} on both the cavity and atoms. The parameters of the system take the same values as in Fig.~\ref{enMK} {\bf (a)} and for $\kappa/2\pi{=}2.5\times10^6$ Hz (corresponding to finesse $F=3\times10^4$). }
\label{wignerm}
\end{figure}

\begin{figure}[t]
\includegraphics[width=0.55\linewidth]{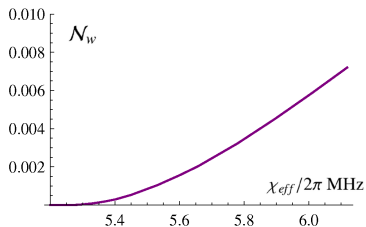}
\caption{Mechanical non-classicality ${\cal N}_w$ versus effective couping $\chi_{e\!f\!f}$ with Geiger-like detection on the atomic state. The parameters take the same values as in Fig.~\ref{enMK} {\bf (a)}.}
\label{Nwchi}
\end{figure}

\section{Non-classicality of the mirror} 
\label{noncalssical}

Having observed the tripartite quantum nonlocality and entanglement in the system, we now turn to the study of quantum effects in its subsystems. We focus on the non-classicality of the motional state of the mechanical system, owing to its significance in the fundamental research in quantum physics~\cite{zurek,collapse}. In what follows, non-classicality is indicated by regions where the Wigner function attains negative values~\cite{wigner}. Due to the linearization of the dynamics and the Gaussian nature of the input states and the noises,  the state of the system is Gaussian at all times, and thus it will not be possible to observe non-classicality simply by tracing out the atomic and cavity subsystems from the joint state. As shown in Ref.~\cite{jiesnjp}, conditional non-Gaussian measurements, e.g. Geiger-like detection, on the cavity may induce a negative Wigner function of the mechanical state. We show in the following that this is also the case for measurements on the atomic mode.
 
We acquired the characteristic function of the system $\zeta(O)$ previously. After the replacements $q={\rm Re}[\alpha], p{=}{\rm Im}[\alpha]; X{=}{\rm Re}[\beta], Y{=}{\rm Im}[\beta]$; and $x{=}{\rm Re}[\gamma], y{=}{\rm Im}[\gamma]$ with amplitude $\{\alpha,\beta,\gamma\}\in\mathbb{C}$, the characteristic function is rewritten as $\zeta(\alpha,\beta,\gamma)$. This gives us access to the density matrix of the system~\cite{glauber}:
\begin{equation}
\rho_{mca}=\frac{1}{\pi^3}\iint \text{d}^2\alpha\,\text{d}^2\beta\,\text{d}^2\gamma\,\zeta(\alpha,\beta,\gamma)D_m(-\alpha)D_c(-\beta)D_a(-\gamma),
\label{rho3}
\end{equation}
where $D_j(\mu)$ is the Weyl operator of mode $j=m,c,a$~\cite{Wallsbook}. Now we implement Geiger-like detection on the cavity and/or the atomic mode. The latter can be carried out by using the quantum jump detection scheme described in Ref.~\cite{Wineland} and employed, e.g., in Ref.~\cite{shelving}. This gives rise to the following density matrix for the conditional mechanical state:
\begin{equation}
\rho_m^{G}={\rm Tr}_{c,a}\left[\Pi^G \, \rho_{mca}\right]/{\rm Tr}_{m,c,a}\left[\Pi^G \, \rho_{mca}\right],
\label{rhomirror}
\end{equation}
where operator $\Pi^G{=}\sum_{s{=}1}^\infty |s\rangle\langle s|$ denotes Geiger-like detection on the cavity/atomic mode ($\Pi^G{=}\sum_{n,m{=}1}^\infty |n\rangle\langle n||m\rangle\langle m|$ for simultaneous detecion on both the subsystems), and the denominator is a normalization constant.

In Fig.~\ref{wignerm}, we present the Wigner distribution of the mirror with Geiger-like detection performed on the system. It has been demonstrated~\cite{jiesnjp} that non-classicality of the mirror can be induced by Geiger-like detection on the cavity field.  Based on the parameters adopted in our system, however, we did not find the negativity of the Wigner function when performing measurements on the cavity. On the contrary, detection on the atoms induces effectively a negative Wigner function [cf. Figs.~\ref{wignerm} {\bf (a)} and {\bf (b)}]. This is probably because, under the parameters, the atoms-mirror coupling or entanglement via the cavity field is stronger than the mirror-light's though there is no direct interaction between them~\cite{vitali3}. For measurements simultaneously performed on both the cavity and the atomic mode, the negativity induced by the detection on the atoms vanishes due to the combined effects caused by the detection on the cavity, as shown in Fig.~\ref{wignerm} {\bf (c)}. The non-classicality is robust against the temperature and the mechanical damping rate~\cite{jiesnjp}: the negativity of Wigner function still survives as the temperature/damping rate increases by three orders of magnitude based on the parameters used in Fig.~\ref{wignerm} {\bf (b)}.

Finally, we show that this measurement-induced non-classicality of the mechanical state is tightly connected with the coupling strength of the system: larger coupling leads to a more nonclassical conditional state, as shown in Fig.~\ref{Nwchi}.  %Here the entanglement is increased by enhancing the optomechanical/atom-light coupling, that is the procedure completed in Fig.~\ref{enMK} {\bf (a)}. 
We have adopted quantity ${\cal N}_w$ to quantify the non-classicality of the state, which is defined as~\cite{wigner,jiewigner} 
\begin{equation}
{\cal N}_w=-\!\int_{\Phi}W_m(\alpha)\text{d}^2\alpha,
\label{eq13}
\end{equation}
where $W_m(\alpha)$ is the Wigner function of the mirror and $\Phi$ is the negative regions of the Wigner distribution in phase space.

\section{Conclusions} 
\label{concl}

We have studied quantum effects in a hybrid optomechanical system by looking at both the tripartite nonlocality and the non-classicality of the mechanical system. The MK inequality is violated demonstrating nonlocal correlations shared among the system. Counterintuitively, the nonlocality shows a negative relation with the tripartite entanglement, in that nonlocality declines as entanglement increases. The negative relation still holds even if bipartite nonlocality is present in the violation of the MK inequality. Our work provides a concrete demonstration in multipartite mixed states of the negative relation between nonlocality and entanglement, and therefore strengthens the link between these two fundamental concepts in quantum physics.

We also studied non-classicality of the motional state of the vibrating mirror. By implementing post-selected measurements, e.g. Geiger-like detection, on the collective atomic mode, a nonclassical mechanical state is generated indicated by the appearance of a negative Wigner function. By enhancing the coupling strength of the system, the mechanical non-classicality increases remarkably. This work predicts the possibility for the experimental realization of nonlocal correlations among atoms, light and a mesoscopic mirror and also contributes to the ongoing attempts of preparing mesoscopic/macroscopic quantum states.

\section{Acknowledgments} 

We are grateful to M. Paternostro and G. Adesso for valuable discussions. %J.L. acknowledges the Institute of Opto-Electronics, Shanxi Univ., and Beijing Computational Science Research Center for hospitality at the early stage of the work. 
This work is supported by National Basic Research Program of China (2012CB921601), National Natural Science Foundation of China (Grants Nos. 11125418, 91336107, 61227902, and 61121064), Natural Science Foundation of Shanxi (Grant No. 2013021005-2), the Royal Commission for the Exhibition of 1851 and the European Commission through the Marie-Curie ITN cQOM.


\begin{thebibliography}{99}
\bibitem{notecoupling} This choice of parameters is enlightened by the work~\cite{jie3ions} that studied tripartite nonlocality of the vibrational modes of three trapped ions. A triangle configuration of the ions, rather than a linear configuration, leads to a higher violation of the MK inequality (${\cal M}_{max}^{Lin}{=}3$, while ${\cal M}_{max}^{Tri}{>}4$). For the triangle configuration the coupling (Coulomb interaction) between any two ions are equal and strongest, while for the linear configuration, the coupling between the two end-chain ions is the weakest.  %Analogously, in this paper we set equal optomechanical and atom-light coupling with the aim to get a higher MK value.
\end{thebibliography}
\end{document}